\documentclass[conference]{IEEEtran}
\usepackage[utf8]{inputenc}
\usepackage[T1]{fontenc}
\usepackage[english]{babel}

\usepackage{textcomp}

\usepackage{graphicx, float, subfigure, blindtext}
\usepackage{physics}
\usepackage{amsmath}
\usepackage{tikz}
\usepackage{mathdots}
\usepackage{yhmath}
\usepackage{cancel}
\usepackage{color}
\usepackage{siunitx}
\usepackage{array}
\usepackage{multirow}
\usepackage{amssymb}
\usepackage{gensymb}
\usepackage{tabularx}
\usepackage{booktabs}

\newcommand\IEEEhyperrefsetup{
bookmarks=true,bookmarksnumbered=true,%
colorlinks=true,linkcolor={black},citecolor={black},urlcolor={black}%
}

\usepackage[\IEEEhyperrefsetup]{hyperref}

\usepackage{mathtools}

\usepackage[nolist,nohyperlinks]{acronym}
\usepackage{cleveref}
\graphicspath{{images/}}
\acrodef{IC}[IC]{Integrated Circuit}
\newacro{svm}[SVM]{Support Vector Machine}

\author{

\IEEEauthorblockN{
Ayush Salik
}

\IEEEauthorblockA{
BINDS Lab, CICS\\
University of Massachusetts, Amherst\\
Email: asalik@umass.edu
}
\and
\IEEEauthorblockN{
Manor Askenazi
}
\IEEEauthorblockA{
Biomedical Hosting LLC\\
Arlington, MA\\
Email: manor@biomedical.hosting
}
\and
\IEEEauthorblockN{
Edward Rietman
}

\IEEEauthorblockA{
BINDS Lab, CICS\\
University of Massachusetts, Amherst\\
and Small-Technology Incubator LLC\\
Grantham, NH\\
Email: erietman@gmail.com
}

} 
\title{Content Addressable Parallel Processors on a FPGA}
\begin{document}
\maketitle
\begin{abstract}
    In this short article, we report on the implementation of a Content Addressable Parallel Processor using a FPGA. While Content addressable memories have been implemented in FPGAs, to our knowledge this is the first implementation in FPGA of Caxton C. Foster's vision of parallel \emph{processing}, particularly the notions of parallel write as well as the combining of output values, which are usually missing in more typical CAM implementations, such as the ones designed for network routing. The resulting CAPP is made accessible to a host computer over a USB/UART interface, using a straightforward serial protocol that is demonstrated using a Python-based driver.
\end{abstract}
\begin{center}
\end{center}
\section{Introduction}
Content Addressable Parallel Processors (CAPPs) constitute an alternative to the standard von Neumann architecture, featuring parallel-processing based on content-addressable memory. Unlike Random Access Memory (RAM) which works by performing operations on a word in memory by referring to its physical address, a CAPP is able to select multiple memory locations simultaneously based on pattern matching the contents of the memory being accessed. Consequently, a CAPP can perform operations like writing into and reading from multiple selected memory cells in constant time. The original intent of the CAPP design was to serve as a general purpose computer, capable of parallel processing. Furthermore, it was hoped that, by providing native support for parallel pattern-matching and multi-write capability, the software for such machines would be relatively easy to write (at least in comparison with other parallel architectures). In practice, this did not occur and CAPPs eventually found use primarily in a much more limited form of application specific CAMs, primarily in the area of computer networking devices, e.g. in the form of fast lookup tables used in network switches and routers.

The goal of our project was to expose a true CAPP capable of multi-write and parallel read, as a convenient USB peripheral, to enable renewed experimentation with this class of architecture. We therefore designed a Verilog module for a parameterized (hence, scalable) CAPP. To this module we added a USB/UART interface module which we manage using an FSM-based protocol. The combined system was implemented on a TinyFPGA-BX \cite{tinyfpga_bx} which we then control over the USB/UART using a simple Python-based driver. We believe that CAPPs have potential in graph theory, neural network caching, high-speed indexing, regex computations etc. and we hope that an open-source, expandable CAPP design which can be synthesized on an FPGA using open-source tools, can provide a basis for renewed research into the application of CAPPs to these diverse and challenging domains. Our implementation is fully open source and can be found here \cite{CAPP_FPGA}.

\section{Content Addressable Parallel Processor}
\begin{figure}
  \includegraphics[height=5cm]{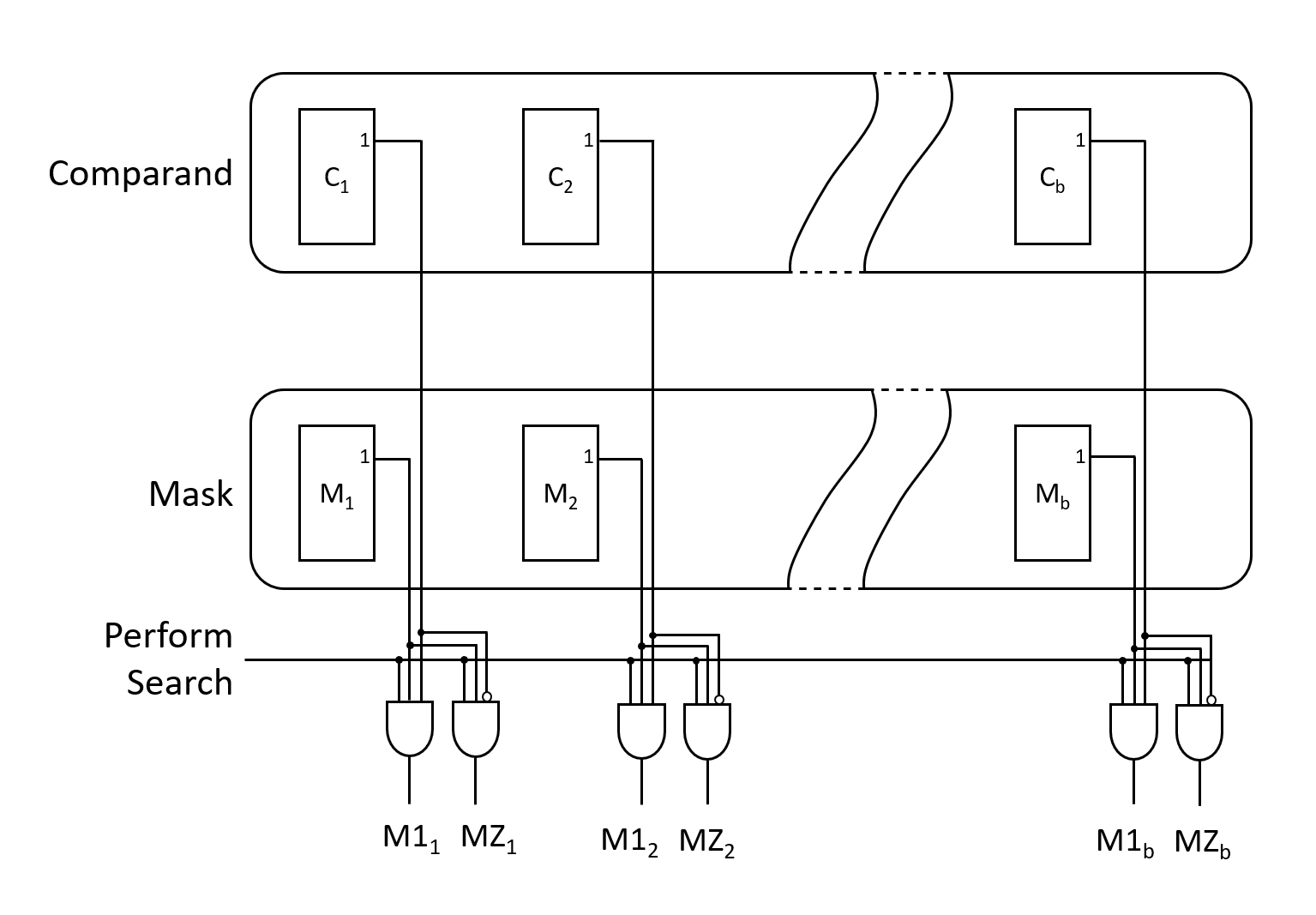}
  \caption{Design of the CAPP search registers}
  \label{search_circuit}
\end{figure}

The circuitry of a CAPP can be broken down into three main parts. The tags, the cells and the search registers. 
The search registers are the least complex circuits in a CAPP. They consist of two different registers, the comparand and the mask. The comparand is the word to search for, and the mask contains locations of bits in the word to ignore during the search. The design for these are given in figure \ref{search_circuit}.

\begin{figure}
  \includegraphics[height=14cm]{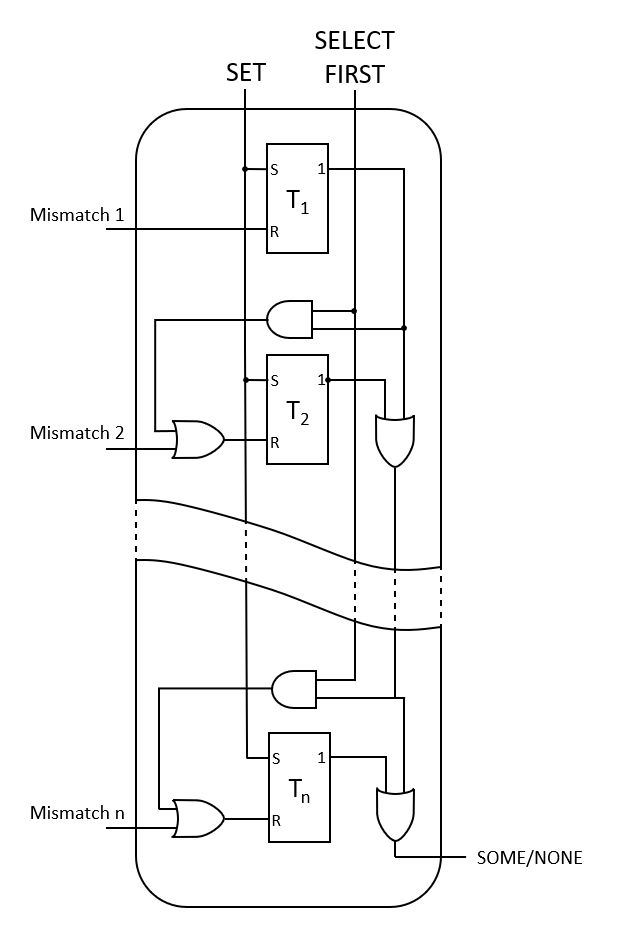}
  \caption{Design of the CAPP tag registers}
  \label{tag_registers}
\end{figure}

The second most complex circuitry is found in the tags. Each tag is a one bit register and there is one tag associated with each memory cell. If the bit of a cell is high after a search, it means that the word in that cell matched the search criteria. In other words, after a search, the cells with high tag bits are the ones that constitute the answer set. As all the tag bits of a CAPP have to change in constant time (i.e. in parallel), match lines from the search registers go through each cell to create mismatch lines which turn irrelevant tags off whenever the search signal is set. The circuitry also includes logic for manually setting all bits high and selecting the first tag bit that is on. The design for these are given in Figure \ref{tag_registers}.

\begin{figure}
  \includegraphics[height=6.5cm]{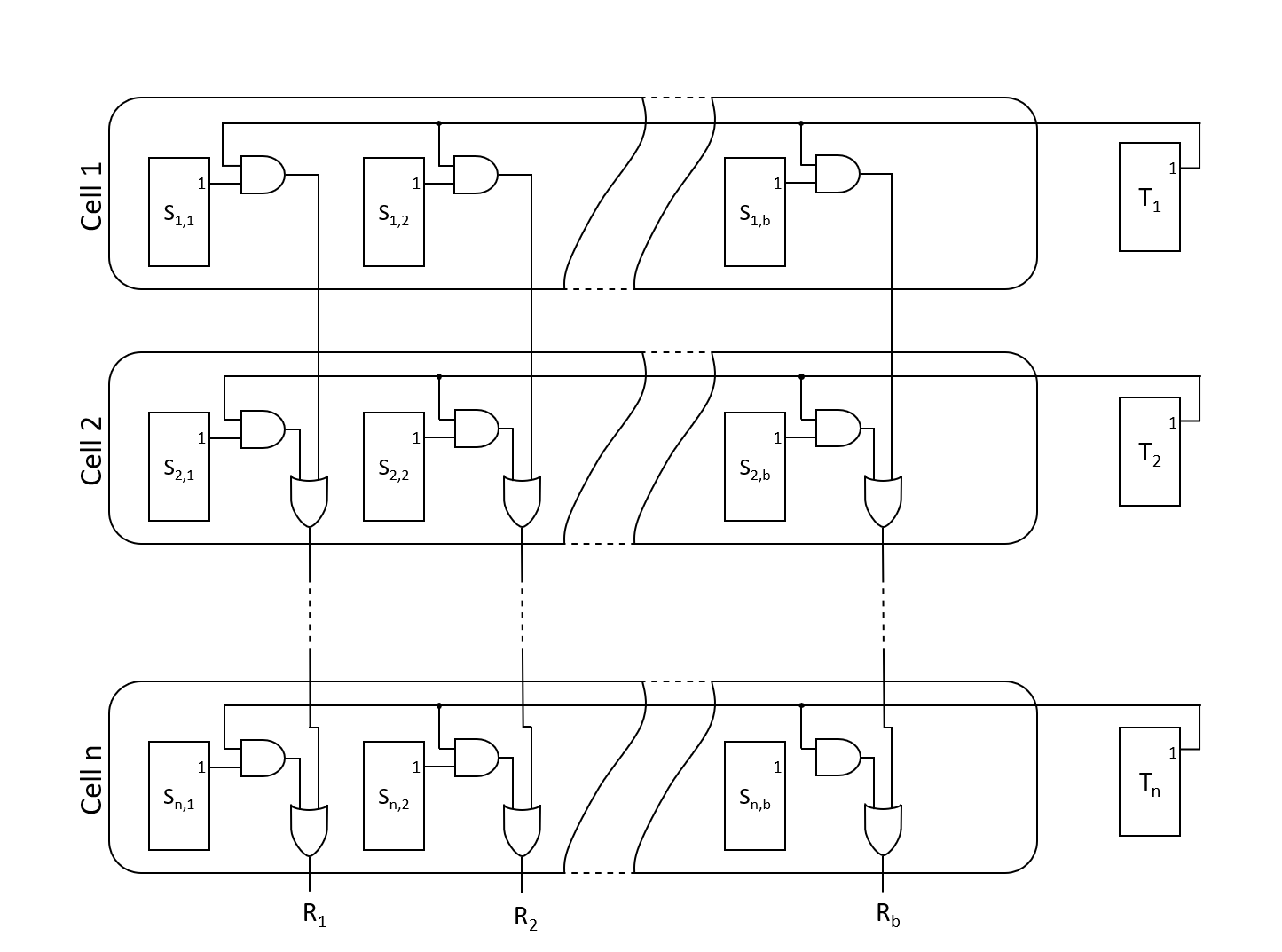}
  \caption{Design of the parallel read logic}
  \label{read_lines}
\end{figure}

\begin{figure}
  \includegraphics[height=6.5cm]{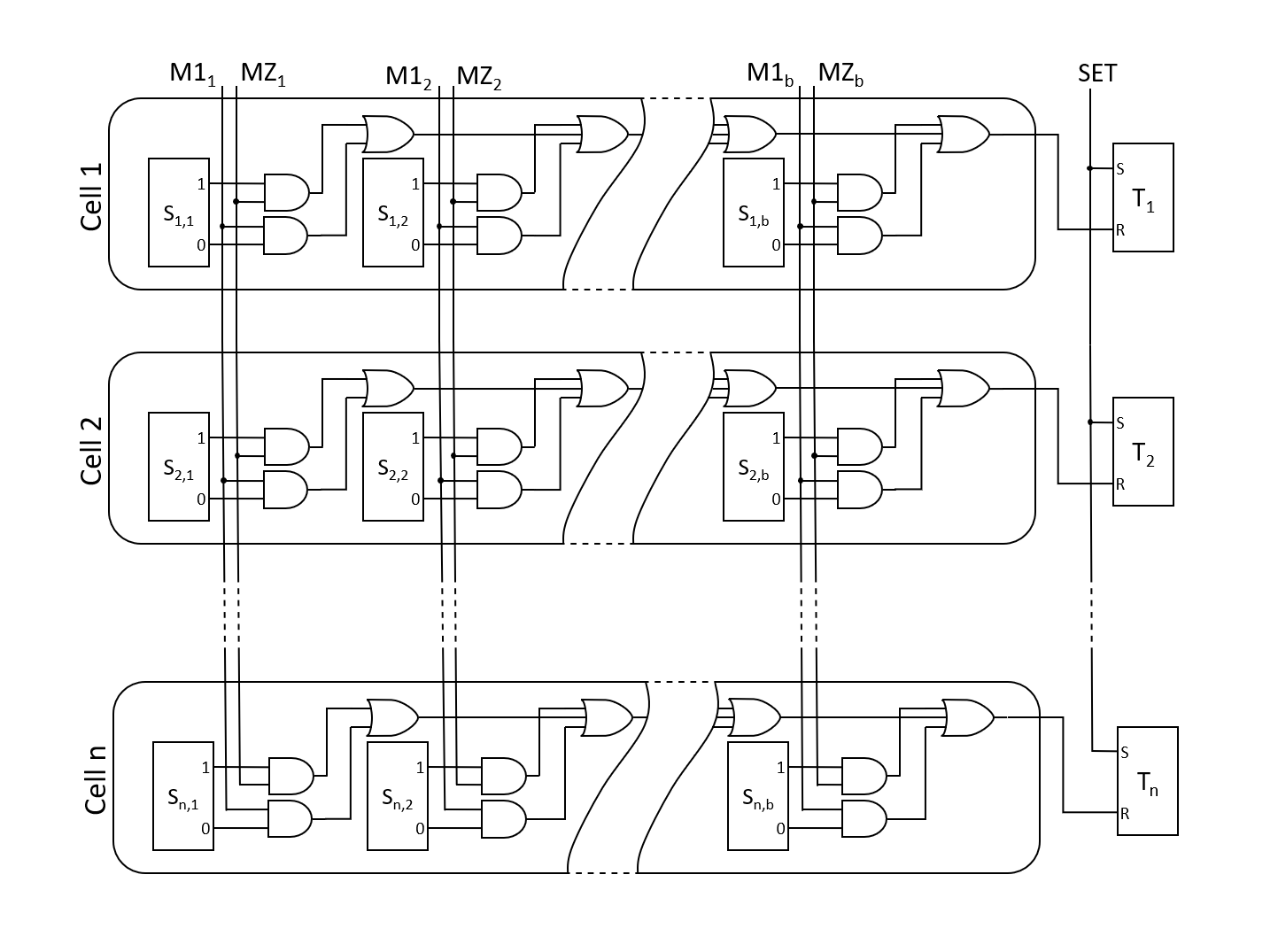}
  \caption{Design of the parallel matching logic}
  \label{mismatch_lines}
\end{figure}

Arguably the most complex circuitry is found in the memory cells themselves, as it encapsulates the main logic for writing, reading and searching. Each bit of each cell is surrounded by two write lines, two search lines and one read line. The n-th bits across all cells share these lines, i.e. each position has a set of lines across all the cells. The value of each bit in each cell is changed to 0 or 1 depending on which write line is set high. Conversely, depending on the two search lines and the bit value, a mismatch line that is connected to the tag bit of the cell is set high or low  (see \ref{mismatch_lines}). This allows the CAPP to search or write in parallel. The output of all the read lines is the bitwise OR of all cells that have their tag bits on as shown in Figure \ref{read_lines}. The parallel writing and reading of results is precisely what distinguishes CAPPs from the more common CAM architecture (as described in the introduction).

The design described above is identical to the introductory CAPP described in Caxton Foster's seminal book on the topic of CAPPs \cite{capp}. We implemented a direct translation to Verilog and then used nextpnr \cite{8735573} for the placing and routing of our Verilog design. The choice of nextpnr was due to its ability to take into account timing constraints which were important for the correct implementation of the Serial UART module needed to function reliably despite being connected to a CAPP module which requires potentially more than one clock cycle to execute a request communicated from the host computer. The resulting layout was synthesized and tested on the TinyFPGA-BX using a Python-based driver to communicate with the resulting USB-CAPP device.
The Finite State Machine at the core of the communication protocol between the host computer and the CAPP is described in the next section.
\section{The Finite State Machine}
The FSM manages the interactions between the CAPP and the host computer and it is responsible for orchestrating the individual CAPP operations necessary to implement complex arbitrary algorithms on the host computer.
It has numerous states, such as those for sending or receiving data, loading data into the CAPP, searching the memory (which in turn requires setting the comparand and mask) as well as reading the combined output. To implement the communication with the host computer we use base UART module provided in David Things' Github repository \cite{uart}. To reduce complexities and minimize our LUT-budget, we adopted the 48MHz clock speed used by the UART module throughout this project. 

\begin{figure}
  \includegraphics[height=6cm]{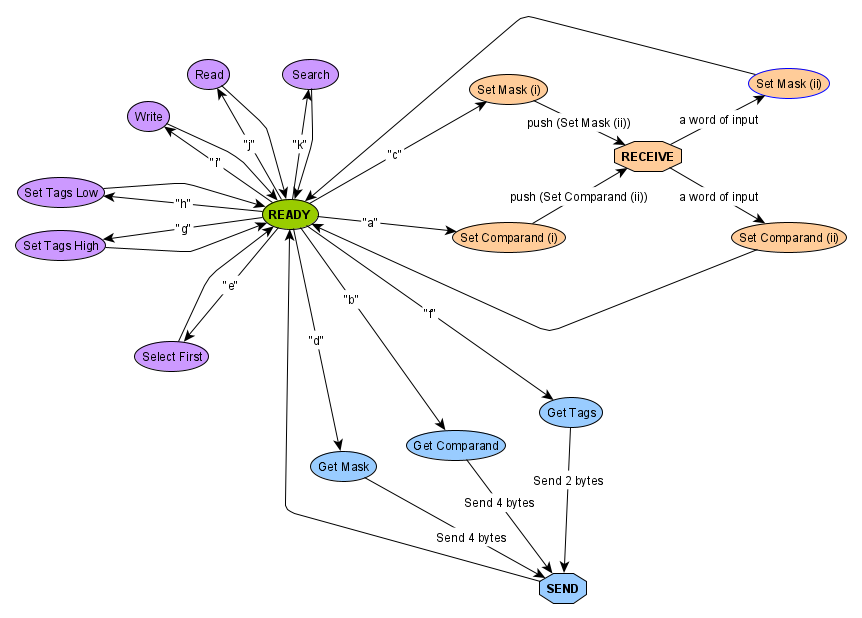}
  \caption{Overview of the FSM underlying the CAPP-host protocol}
  \label{FSM_protocol}
\end{figure}

As shown in Figure \ref{FSM_protocol}, our FSM will typically break down a task into multiple short procedural tasks to reduce both magnitude and variance of the latency associated with each atomic state/task. In other words, we attempted to have a repertoire of tasks which all had (similarly) short latencies. Consequently, an overall task may necessitate several atomic states before returning the CAPP to its default ready state. 
For example, the operations for searching involve several steps, which are typically initiated by the Python "driver" and include:

\begin{itemize}
    \item Resetting the tags (set tags high, followed by set tags low)
    \item Setting the comparand 
    \item Setting the mask 
    \item Sending the search signal 
\end{itemize}

Notice that setting the comparand and mask require several clock cycles as the memory is based on 4-byte words and the CAPP can only send one byte per CAPP clock cycles. Furthermore, the FSM diagram shows a single RECEIVE state that can be used to receive both the comparand and the mask. Technically, in a formal FSM diagram, this should be represented by two distinct states, however, the Verilog code re-uses some logic when implementing these operations and we wanted the FSM-diagram to reflect this.

As seen above, resetting the tags is, in fact, a two step operations:

\begin{itemize}
    \item Set Tags High: change SET to high
    \item Set Tags Low: change SET to low
\end{itemize}
where SET refers to the SET line shown in Figure \ref{tag_registers}.

Some operations that appear to be atomic in Figure \ref{FSM_protocol} are in fact composed of multiple micro-steps, including an "IDLE" (delay) step. For example, the Search command is in fact composed of three micro-steps:

\begin{itemize}
    \item SEARCH\_1: 
        \begin{itemize}
            \item set SEARCH to high
            \item set delay to 5 clock cycles
        \end{itemize}
    \item IDLE:
        \begin{itemize}
            \item wait for duration of delay
            \item go to SEARCH\_2
        \end{itemize}
    \item SEARCH\_2: 
        \begin{itemize}
            \item set SEARCH to low
        \end{itemize}
\end{itemize}

where SEARCH refers to the "Perform Search" line in Figure \ref{search_circuit}. IDLE represents another case of a "shared" micro-step which is reused by two commands (as was the case with RECEIVE). Though it is not represented in Figure \ref{FSM_protocol} (in the interest of clarity), IDLE is in fact used by both the Search and Select commands.

\section{Results}
The Content Addressable Parallel Processor was successfully implemented on a TinyFPGA-BX. It was tested both by manual interaction with the serial USB-UART as well as through a python driver which wraps the protocol in a higher-level object-oriented API. The python driver file also contains a simple test program that serves both as a tutorial and a test-suite for the FPGA based CAPP.
\section{Discussion}
Historically, CAPPs have clearly lost out to von Neumann machines. Arguably, this was because the increased price associated with the more complex memory cells was too prohibitive, especially in the early days of computing. However, over the years, manufacturing and designing costs have plummeted while the difficulty of writing correct parallel software has remains as high as it ever was. This makes it a good time to reconsider alternative architectures, which, by exposing powerful primitives such as content addressable read and write operations should simplify the development of parallel algorithms. We have therefore revived Caxton Foster's CAPP research project in the form of an FPGA-based co-processor, which we think will be able to support new, interesting and competitive implementation of algorithms in areas such as: fast lookup tables, neural networks, parallel regex operations, graph algorithms and more. Our implementation is fully open source and can be found here \cite{CAPP_FPGA}.
\section*{Acknowledgment}
We acknowledge Kartik Thakore of Sharecare Inc. for technical feedback. ER was funded by a contract from Doc.AI Inc. MA was funded by a contract from Small-Technology Incubator LLC. PowerPoint templates of basic Logic Gates were designed by Oliver Mannay and downloaded from: \url{https://www.tes.com/en-us/teaching-resource/logic-gate-symbols-for-powerpoint-11139006}
\bibliographystyle{IEEEtran}
\bibliography{IEEEabrv,references}
\end{document}